\documentclass[pra]{revtex4}
\usepackage{amssymb}
\usepackage{amsmath}
\usepackage{graphicx}
\usepackage{dcolumn}
\usepackage[center]{subfigure}
\usepackage{color}
\begin{document}

\title{Geometric phase with full-wedge and half-wedge rotation in nonlinear frequency conversion}

\author{Feiyan Zhao$^{1}$, Jiantao L\"{u}$^{1,2}$, Hexiang He$^{1,2}$, Yangui Zhou$^{1,2}$, Shenhe Fu$^{3}$, Yongyao Li$^{1,2}$}
\email{yongyaoli@gmail.com}
\affiliation{$^{1}$ School of Physics and Optoelectronic Engineering, Foshan University, Foshan 528000, China \\
$^{2}$Guangdong-Hong Kong-Macao Joint Laboratory for Intelligent Micro-Nano Optoelectronic Technology, Foshan University, Foshan 528000, China \\
$^{3}$Department of Optoelectronic Engineering, Jinan University, Guangzhou 510632, China
}

\begin{abstract}
When the quasi-phase matching (QPM) parameters of the $\chi^{(2)}$ nonlinear crystal rotate along a closed path, geometric phase will be generated in the signal and idler waves that participate in the nonlinear frequency conversion. In this paper, we study two rotation schemes, full-wedge rotation, and half-wedge rotation, of the QPM parameters in the process of fully nonlinear three-wave mixing. These two schemes can effectively suppress the uncertainty in creating the geometric phase in the nonlinear frequency conversion process when the intensity of the pump is depleted. The finding of this paper provides an avenue toward constant control of the geometric phase in nonlinear optics applications and quantum information processing.
\end{abstract}

\maketitle

\section{Introduction}

When the parameters of a quantum system are slowly cycled around a closed path, a geometric phase is acquired in addition to the dynamical phase for the quantum state \cite{1:Berry}. In the past 30 years, the geometric phase has been shown to have broad applications in many areas of physics, including quantum technology \cite{2:wubiao,3:Mart,4:Leek,5:Jxu}, cold atom physics \cite{6:Shengchang,7:Shengchang}, condensed-matter physics \cite{8:Dxiao}, and optics \cite{9:Berry}. In optics, the geometric phase commonly refers to the Pancharatnam-Berry (PB) phase, which usually applies to light propagating through an optical medium with slowly changing polarization or direction of photons. It is widely used for beam shaping \cite{10:Bomzon,11:Marrucci}, shearing \cite{12:AAC}, guiding \cite{13:Slussarenko}, trapping \cite{14:Jisha}, holography \cite{15:Tzhan}, and encoding of classical \cite{16:Milione} or quantum \cite{17:Stav} information. Very recently, a photonic spin-Hall effect was created by a wavevector-varying PB phase \cite{18:WZhu}.

Another manifestation of the geometric phase in optics is its extension to the nonlinear optics region. One extension uses circularly polarized light on a nonlinear metasurface \cite{19:Tymchenko,20:GuixingLi,21:GuixinLi,22:Devlin,23:ZLiang}. Another involves accumulation of the geometric phase in a bulk material during the frequency conversion along a nonlinear crystal or waveguide through the quasi-phase-matching (QPM) technique \cite{24:Kwang}. Recently, in terms of the latter case, Karnieli and Arie found a controllable geometric phase in the process of sum-frequency generation (SFG) with an undepleted pump through $\chi^{(2)}$ nonlinear crystals by analogy between the SFG and the spin-1/2 system \cite{25:Aviv}. The geometric phase in this case can be well predicted via the available linear theory on the dynamics of a two-level system. Then, they experimentally realized nonreciprocal transmission and wavefront shaping through the geometric phase using such an SFG process \cite{26:Aviv}. They also used this process to emulate high-order skyrmions and the topological Hall effect in nonlinear optics \cite{27:Aviv}. However, if the pump wave becomes comparable to that of the idler or signal, then the undepleted pump approximation cannot hold, which gives rise to invalidation of the superposition principle and the conservation of the total photon number. Under this condition, the linear theory in the spin-1/2 or two-level system cannot be applied to calculate the geometric phase. The calculation of the geometric phase in a nonlinear theory remains an open question \cite{28:Garrison,29:Alber,30:JieLiu}. Very recently, a theoretical analysis developed from the geometric representation of three-wave mixing has provided a method for precisely calculating the geometric phase when the pump wave becomes depleted \cite{31:Yongyao}. This method is universal for both undepleted and depleted pump cases and has been extended to the calculation of the geometric phase in the four-wave mixing process \cite{32:Yongyao}. However, this method has been applied only to circular rotation of the QPM parameter thus far. The magnitude of the geometric phase deviates from that of the theoretical prediction in the undepleted pump case if the pump wave becomes depleted. This deviation is nonmonotonic, which makes practical application difficult if one wants to build a pump-controlled device using the geometric phase. Hence, realizing how to acquire a constant geometric phase with different pump depletion levels or how to let the geometric phase have a well-monotonic and well-predicted behavior in the undepleted pump case becomes necessary for developing an effective device for the pump-controlled phase modulator.

\begin{figure}[ht!]
\centering\includegraphics[width=0.8\columnwidth]{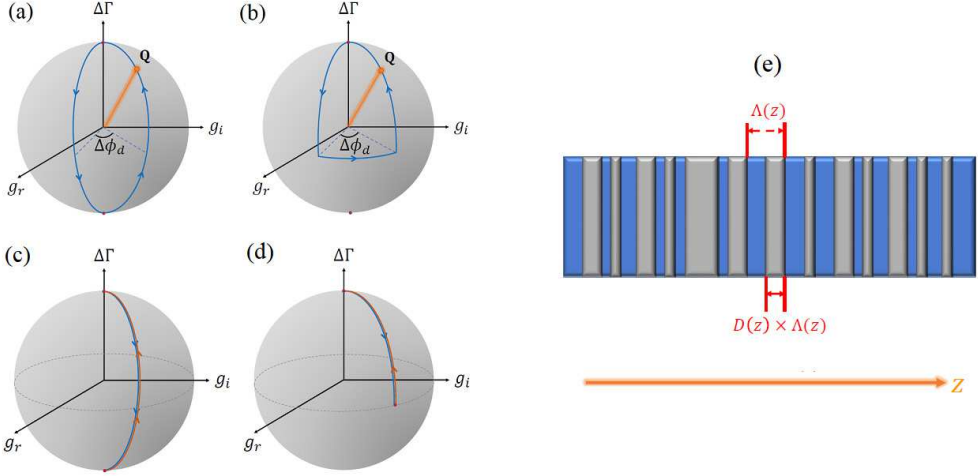}
\caption{(a,b)Scheme of full-wedge rotation and half-wedge rotation of the vector $\mathbf{Q}$ in the parameter space, respectively. In this paper, the integral along these two trajectories are represented by the symbol '$\bigcirc$'. (c,d)Trajectories of the round-trip motion corresponding to full-wedge and half-wedge rotation in Eq. (\ref{geometric4}) (the symbol of the integral is `$\downharpoonleft\upharpoonright$'). The blue and the red parts in the trajectory are referred to in the `$\downharpoonleft$' and `$\upharpoonright$' parts, respectively. (e)Sketch of the modulation of the nonlinear crystal. The local period $\Lambda(z)$ adds a modulation wavevector $K_{\Lambda}(z)=2\pi/\Lambda(z)$. Each domain width of the opposite poling of the crystal is represented by the duty cycle as $D(z)\times\Lambda(z)$, and its relative position within the periodic cycle is defined by a factor, $\phi_{\mathrm{d}}$.} \label{sketch}
\end{figure}

In this paper, we use full-wedge rotation of the QPM vector in the parameter space [see Fig. \ref{sketch}(a)] to design the nonlinear modulation pattern along nonlinear crystals. In contrast to the previous study of circular rotation, we find that the magnitude of the geometric phase in the SFG and difference-frequency generation (DFG) processes becomes independent of the pump depletion level. By comparison with circular rotation, we find that this effect may be due to the nonadiabatic variation in the phase factor of the QPM parameter in the wedge rotation. To demonstrate this, we further design half-wedge rotation [see \ref{sketch}(b)], in which all QPM parameters undergo adiabatic rotation. We find that the geometric phase becomes dependent on the pump depletion level again. Hence, the adiabatic or nonadiabatic variation in the QPM phase factor plays an important role in the creation of the geometric phase when the pump becomes depletion. Moreover, the geometric phase in the half-wedge rotation process manifests a well-monotonic and well-predicted behavior, which is proportional to the wedge angle. Ref. \cite{33:Jiantao} also tried to produce a constant geometric phase in the depleted pump case for the circular rotation. However, unlike the results in the current paper, the constant geometric phase in Ref. \cite{33:Jiantao} requires a very specifical modulation pattern for the nonlinear crystal with a specifical requirement in the initial condition. These restricted conditions strongly limit its application.
The rest of the paper is structured as follows. The theory of the method for calculating the geometric phase in the depleted pump case is briefly illustrated in Sec. II. The geometric phases in the full-wedge rotation and half-wedge rotation are discussed in Sec. III. In Sec. IV, we provide a potential application of utilizing the geometric phase for all-optical beam shaping in the case of second-harmonic generation (SHG)---i.e., a fully depleted pump case. The main results of the paper are summarized in Sec. V.

\section{Model and theory}

\subsection{Dynamical equation for the three-wave mixing process}
The evolution of the slowly varying envelopes of the electric fields in the three-wave mixing process
in the QPM grating is governed by the coupled quasi-CW equations (CWEs) \cite{34:Suchowski,35:Gil}
\begin{eqnarray}
&&\partial_{z}A_{1}=-i\frac{2d(z)\omega_{1}}{c n_{1}}A^{\ast}_{2}A_{3}e^{-i\Delta k_0 z},\label{Basieq1}\\
&&\partial_{z}A_{2}=-i\frac{2d(z)\omega_{2}}{c n_{2}}A^{\ast}_{1}A_{3}e^{-i\Delta k_0 z},\label{Basieq2}\\
&&\partial_{z}A_{3}=-i\frac{2d(z)\omega_{3}}{ c n_{3}}A_{1}A_{2}e^{i\Delta k_0 z}, \label{Basieq3}
\end{eqnarray}
where $A_{1,2,3}$ are the slowly varying envelopes of the idler, pump, and signal waves, respectively; $\Delta k_{0}=k_{1}+k_{2}-k_{3}$ is the phase mismatch; and $d(z)$ is the spatially varying magnitude of the second-order nonlinear susceptibility, which can be expanded by the Fourier series with slowly varying components as
\begin{eqnarray}
&&d(z)=d_{ij}\sum^{\infty}_{m=-\infty}|d_{m}(z)|\times\notag\\
&&\exp\left\{im\left[\int^{z}_{0}K_{\Lambda}(z')dz'+\phi_{\mathrm{d}}(z)\right]\right\}, \label{Dz1}
\end{eqnarray}
with
\begin{eqnarray}
d_{m}(z)=
\begin{cases}
(2/ m\pi)\sin\left[m\pi D(z)\right] & m\neq0 \\
2D(z)-1 & m=0
\end{cases},\label{Dz2}
\end{eqnarray}
where $d_{ij}$ is the corresponding nonlinear susceptibility tensor. $K_{\Lambda}(z)$, $0<D(z)<1$ and $\phi_{d}(z)$ are three QPM parameters \cite{25:Aviv}. The detail meaning of these three QPM parameters are explained as follows: (i) The local period $\Lambda(z)$ adds a modulation wavevector $K_\Lambda(z) = 2\pi/\Lambda(z)$. (ii) Each domain width of the oppositely poling of the crystal is represented by the duty cycle as $D(z)\times\Lambda(z)$. (iii)The relative position within the periodic cycle is defined by a phase factor $\phi_{\mathrm{d}}$. A sketch map for these 3 QPM parameters is displayed in Fig. \ref{sketch}(e). In Eq. (\ref{Dz1}), because only the $m=\pm1$ terms can provide the best compensation for the phase mismatch, other terms can be dropped from the expansion. The rescaling definitions are applied as \cite{31:Yongyao}
\begin{eqnarray}
&&A_{j}=q_{j}\left(\frac{\omega_{j}}{ n_{j}}\sum^{3}_{l=1}\frac{n_{l}}{\omega_{l}}|A_{l0}|^{2}\right)^{1/2}\times e^{-i\left[\Delta k_{0}z-\int^{z}_{0}K_{\Lambda}(z')dz'\right]},\notag\\
&&\eta=\frac{4d_{ij}}{\pi c}\left(\frac{\omega_{1}\omega_{2}\omega_{3}}{ n_{1}n_{2}n_{3}}\sum^{3}_{l=1}\frac{n_{l}}{\omega_{l}}|A_{l0}|^{2}\right)^{1/2},\notag\\
&&\tau=\eta z,\notag\\
&&\Delta\Gamma(\tau)=\left[\Delta k_{0}-K_{\Lambda}(\tau)\right]/\eta,\notag\\
&&g(\tau)=\sin\left[\pi D(\tau)\right]e^{i\phi_{d}(\tau)}=\Xi(\tau)e^{i\phi_{d}(\tau)}, \label{Acharacters}
\end{eqnarray}
where $A_{l0}$ is the amplitude of wave $\omega_{l}$ at $z=0$. Eqs. (\ref{Basieq1}-\ref{Basieq3}) can be simplified to dimensionless forms as
\begin{eqnarray}
&&\frac{d}{d{\tau}}q_{1}=i\Delta\Gamma q_{1}-igq^{\ast}_{2}q_{3},\notag\\
&&\frac{d}{d{\tau}}q_{2}=i\Delta\Gamma q_{2}-igq^{\ast}_{1}q_{3},\notag\\
&&\frac{d}{d{\tau}}q_{3}=i\Delta\Gamma q_{3}-ig^{\ast}q_{1}q_{2}, \label{Basiceq}
\end{eqnarray}
where $q_{1,2,3}$ are the amplitudes of the photon flux of the three waves; $\Delta\Gamma$ is one of the QPM parameters, which represents the scaled form of the phase mismatch; and $g=\Xi e^{i\phi_{d}}$ is the nonlinear coupling coefficient, where $\Xi$ and $\phi_{d}$ correspond to the other two QPM parameters, which represent the duty cycle and the phase of the modulation in the nonlinear crystals, respectively. Here, we use the $m=-1$ term in Eq. (\ref{Dz1}) for coupled Eqs. (\ref{Basieq1},\ref{Basieq2}) and the $m=1$ term for coupled Eq. (\ref{Basieq3}).

The variation in these three QPM parameters can be denoted by the motion of a vector $\mathbf{Q}=(\Xi\cos\phi_{d},\Xi\sin\phi_{d}, \Delta\Gamma/2)$ in the parameter space [See Figs. \ref{sketch}(a,b)]. When the intensity of the pump wave is strong enough, which can be termed as an undepleted wave, the system can be represented by a spin-1/2 system, and the $\mathbf{Q}$ vector can be equivalent to the time-dependent magnetic field \cite{25:Aviv}. The initial conditions for the three waves are defined by the initial intensities of the three waves; i.e., $I_{j}=|q^{(0)}_{j}|^2$. According to the rescaling definition in Eq. (\ref{Acharacters}), the sum of $I_{j}$ automatically satisfies
\begin{eqnarray}
I_{1}+I_{2}+I_{3}=1.
\end{eqnarray}

\subsection{Geometrical representation and calculation of the geometric phase}
To show the geometric motion of Eqs. (\ref{Basiceq}) by means of a Bloch surface in the state space, a set of coordinates $(X, Y, Z)$ and a closed surface $\varphi(X,Y,Z)=0$ are introduced by defining \cite{36:Luther,37:Phillips}
\begin{eqnarray}
&&X+iY=q_{1}q_{2}q^{\ast}_{3},\quad Z=|q_{3}|^2, \label{coordinate}\\
&&\varphi=X^2+Y^2-Z(Z-K_{1})(Z-K_{3}), \label{Sphere}
\end{eqnarray}
where $K_{1}=|q_{1}|^2+|q_{3}|^{2}$ and $K_{3}=|q_{2}|^2+|q_{3}|^2$ are constants, which are borrowed from the well-known Manley-Rowe relations in the three-wave mixing process. Hence, all states, which can be defined by a state vector $\mathbf{W}=X\mathbf{\hat{i}}+Y\mathbf{\hat{j}}+Z\mathbf{\hat{k}}$, are mapped on the the Bloch surface $\varphi$.

According to Ref. \cite{31:Yongyao}, the adiabatic geometric phase for the three waves at different pump depletion levels can be calculated via the formula
\begin{eqnarray}
\beta_{j}=\int_{\bigcirc} p^{\prime}_{j}dq^{\prime}_{j}-\int_{\downharpoonleft\upharpoonright}p^{\prime}_{j}d q^{\prime}_{j}=\Phi_{j}-D_{j}, \label{geometric4}
\end{eqnarray}
where $q^{\prime}_{j}=q_{j}/\sqrt{N_{j}}$ and $p^{\prime}_{j}=-iq_{j}/N_{j}$, with $N_{j}=\sqrt{2|q_{j}|^{2}}$. `$\bigcirc$' represents the state vector following the motion of $\mathbf{Q}$ governed by all QPM parameters, $\Delta\Gamma$, $\Xi$, and $\phi_d$, and acquires the total phase $\Phi_{j}$. However, `$\upharpoonleft\downharpoonright$' represents the motion of $\mathbf{Q}$ governed only by two QPM parameters, $\Delta\Gamma$ and $\Xi$ (the third QPM parameter $\phi_d$ is not involved in this motion); the motion is pinned at the same meridian and is allowed to move only up and down along this meridian [see the sketches of this motion in Figs. \ref{sketch}(c,d)]. Hence, we term the `$\upharpoonleft\downharpoonright$' motion as `round-trip' motion. This motion can be viewed as the vertical projection of the `$\bigcirc$' motion and had been demonstrated to acquire only the dynamical phase $D_{j}$ \cite{31:Yongyao}.

\subsection{Full-wedge and half-wedge rotation}

In this paper, we consider the vector $\mathbf{Q}$ to perform full-wedge rotation and half-wedge rotation in the parameter space [see Figs. \ref{sketch}(a) and (b)]. Then, we calculate the geometric phase for these two rotation types.

The full-wedge rotation of $\mathbf{Q}$ is defined as \cite{25:Aviv}
\begin{equation}
\begin{cases}
\Delta\Gamma/2=\cos(\Omega\tau)\\
\Xi=|\sin(\Omega\tau)| \\
\phi_{d}=\left[H(\tau-T/2)-1/2\right]\Delta\phi_d,
\end{cases}
\label{FWedge}
\end{equation}

where $H(\tau-T/2)$ is the Heaviside step function. $\Omega=2\pi/T$ is the rotational speed, $T$ is the effective length of the nonlinear crystals, and $\Delta\phi_{d}$ is the wedge angle. A sketch for this motion is shown in Fig. \ref{sketch}(a), and the corresponding round-trip motion is shown in Fig. \ref{sketch}(c).

For the half-wedge rotation, the motion of vector $\mathbf{Q}$ is divided into three parts: (i) the vector moves from the north pole toward the equator via a meridian of the sphere; (ii) the vector moves via the equator from the original meridian to another meridian with an angle of $\Delta\phi_d$; and (iii) the vector moves back from the equator to the north pole via the meridian at which the second part terminates. Hence, we can divide the effective length of the crystals into three parts: $0\leq\tau\leq T/5$, $T/5<\tau\leq 4T/5$ and $4T/5<\tau\leq T$. The variations in the QPM parameters in these three parts are defined as follows:
\begin{eqnarray}
\begin{cases}
\Delta\Gamma/2=\cos(\Omega'\tau)&  \\
\Xi=|\sin(\Omega'\tau)| & 0\leq\tau\leq T/5 \\
\phi_d=-\Delta\phi_d/2 &
\end{cases},
\end{eqnarray}

\begin{eqnarray}
\begin{cases}
\Delta\Gamma/2=0 &  \\
\Xi=1 & T/5<\tau\leq 4T/5 \\
\phi_d=\left[(\tau-\tau_0)/T_e-1/2\right]\Delta\phi_d &
\end{cases},
\end{eqnarray}

and

\begin{eqnarray}
\begin{cases}
\Delta\Gamma/2=\sin(\Omega'\tau) & \\
\Xi=|\cos(\Omega'\tau)| & 4T/5<\tau\leq T \\
\phi_d=\Delta\phi_d/2 &
\end{cases}
\end{eqnarray}
where $\Omega'=\pi/(2T/5)$ is the angular speed of the vector when moving along the meridian, $\Delta\phi_{d}$ is the wedge angle, $\tau_0=T/5$, and $T_e=3T/5$. A sketch for this motion is shown in Fig. \ref{sketch}(b), and the corresponding round-trip motion is shown in Fig. \ref{sketch}(d).

\begin{figure}[ht!]
\centering\includegraphics[width=0.66\columnwidth]{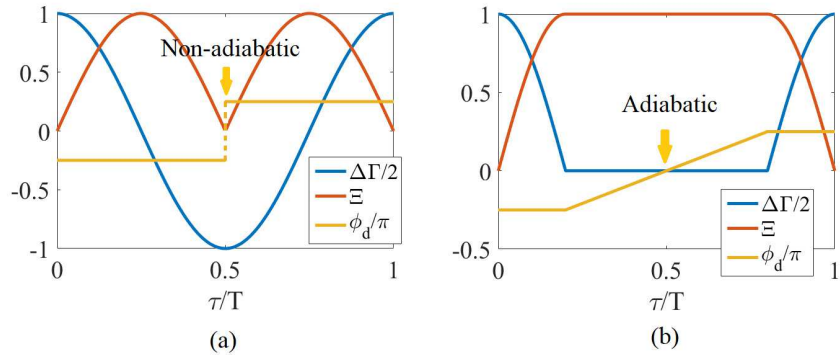}
\caption{Examples of modulation of the QPM parameters in full-wedge rotation (a) and half-wedge rotation (b) with $\Delta\phi_d=\pi/2$. In panel (a), the variation in $\phi_d$ is a nonadiabatic process, while in panel (b), it is an adiabatic process.} \label{Exp2}
\end{figure}

Fig. \ref{Exp2} shows two examples of QPM modulation for these two motions. In the full-wedge rotation process [see the example in Fig. \ref{Exp2}(a)], two of the QPM parameters, $\Delta\Gamma$ and $\Xi$, are described by a continuous function, while the last parameter, $\phi_d$, is described by a step function, which is discontinued at $\tau=T/2$ [see the vertical dashed line in Fig. \ref{Exp2}(a)]. The adiabatic condition for the variation in the three QPM parameters can be roughly estimated by \cite{34:Suchowski}
\begin{eqnarray}
\frac{1}{T}\left|\frac{dQ_{i}}{d\tau}\right|\ll1, \label{adia_condition}
\end{eqnarray}
where $(Q_{1},Q_{2},Q_{3})=(\Delta\Gamma,\Xi,\phi_{\mathrm{d}})$ are the three parameters. Because the current two rotations are defined by piecewise functions. The above condition should be better represented by
\begin{eqnarray}
\frac{1}{T}\left|\lim_{\Delta\tau\rightarrow0}\frac{\Delta Q_{i}}{\Delta\tau}\right|\ll1, \label{adia_condition2}
\end{eqnarray}
If $T$ (the effective length of the crystal) is sufficiently large (in the numerical simulation, we select $T=500$), this condition can be held if the curve of $Q_{i}(\tau)$ is unbroken. In Fig. \ref{Exp2}(a,b), except for the case of $\phi_{\mathbf{d}}(\tau)$ at $\tau=T/2$ in the process of full-wedged rotation, all the QPM curves are unbroken. The $\phi_{\mathrm{d}}(\tau)$ for the case of full-wedged rotation at $\tau=T/2$ satisfies,
\begin{eqnarray}
\lim_{\Delta\tau\rightarrow0}\frac{\Delta\phi_{\mathrm{d}}}{\Delta\tau}\Big|_{\tau=T/2}\rightarrow\infty,
\end{eqnarray}
where $\Delta\phi_{\mathrm{d}}$ is the wedged angle, which is a finite number. Hence, $\phi_{\mathrm{d}}$ undergoes a nonadiabatic variation at this point. However, this nonadiabatic variation does not appear in the trajectory of the $\mathbf{Q}$-vector on the parameter surface. It occurs only at the south pole of the parameter surface [see Fig. \ref{sketch}(a)]. Under linear conditions, such a nonadiabatic variation is always neglected. In contrast, in the half-wedge rotation process [see the example in Fig. \ref{Exp2}(b)], according to the condition in Eq. (\ref{adia_condition2}), all three QPM parameters, $\Delta\Gamma$, $\Xi$ and $\phi_{d}$, vary adiabatically, which is different from the full-wedge rotation case. When the pump wave is undepleted, the geometric phase in these two processes can be predicted by the same theory in spin-1/2 systems. However, once the pump wave becomes depleted, the geometric phases in these two processes show different behavior. In the full-wedge rotation process, the geometric phase is independent of the pump depletion level. By contrast, in the half-wedge rotation process, the geometric phase becomes dependent on the pump depletion level. In the next section, we will provide a systematic discussion of the geometric phases in these two processes.

\section{Numerical simulation}

\subsection{Full-wedge rotation}
\begin{figure}[ht!]
\centering\includegraphics[width=0.99\columnwidth]{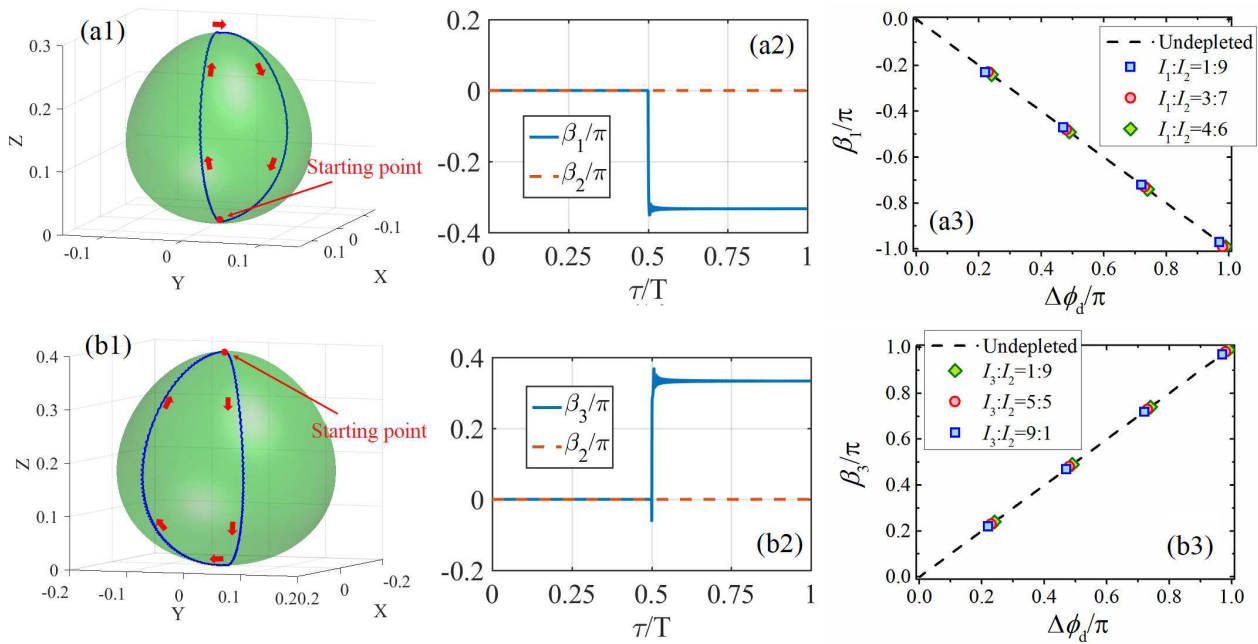}
\caption{(a1) Trajectory of clockwise full-wedge rotation in the SFG process on the Bloch sphere with $I_{1}:I_{2}=4:6$ and $\Delta\phi_{d}=\pi/3$. This process starts from the bottom of the Bloch surface. (a2) Geometric phase accumulation for wave $q_{1,2}$ along $\tau$ with $I_{1}:I_{2}=4:6$ and $\Delta\phi_{d}=\pi/3$. (a3) Geometric phase $\beta_{1}$ versus $\Delta\phi_d$ for different depletion levels. At the undepleted limit, $\beta_{1}=-\Delta\phi_d$ (the black dashed line). (b1) Trajectory of clockwise full-wedge rotation in the DFG process on the Bloch sphere with $I_{3}:I_{2}=4:6$ and $\Delta\phi_{d}=\pi/3$. This process starts from the top of the Bloch surface. (b2) Geometric phase accumulation for wave $q_{3,2}$ along $\tau$ with $I_{3}:I_{2}=4:6$ and $\Delta\phi_{d}=\pi/3$. (b3) Geometric phase $\beta_{3}$ versus $\Delta\phi_d$ for different depletion levels. At the undepleted limit, $\beta_{3}=\Delta\phi_d$ (the black dashed line). } \label{Fwedgeresult}
\end{figure}

\begin{figure}[ht!]
\centering\includegraphics[width=0.99\columnwidth]{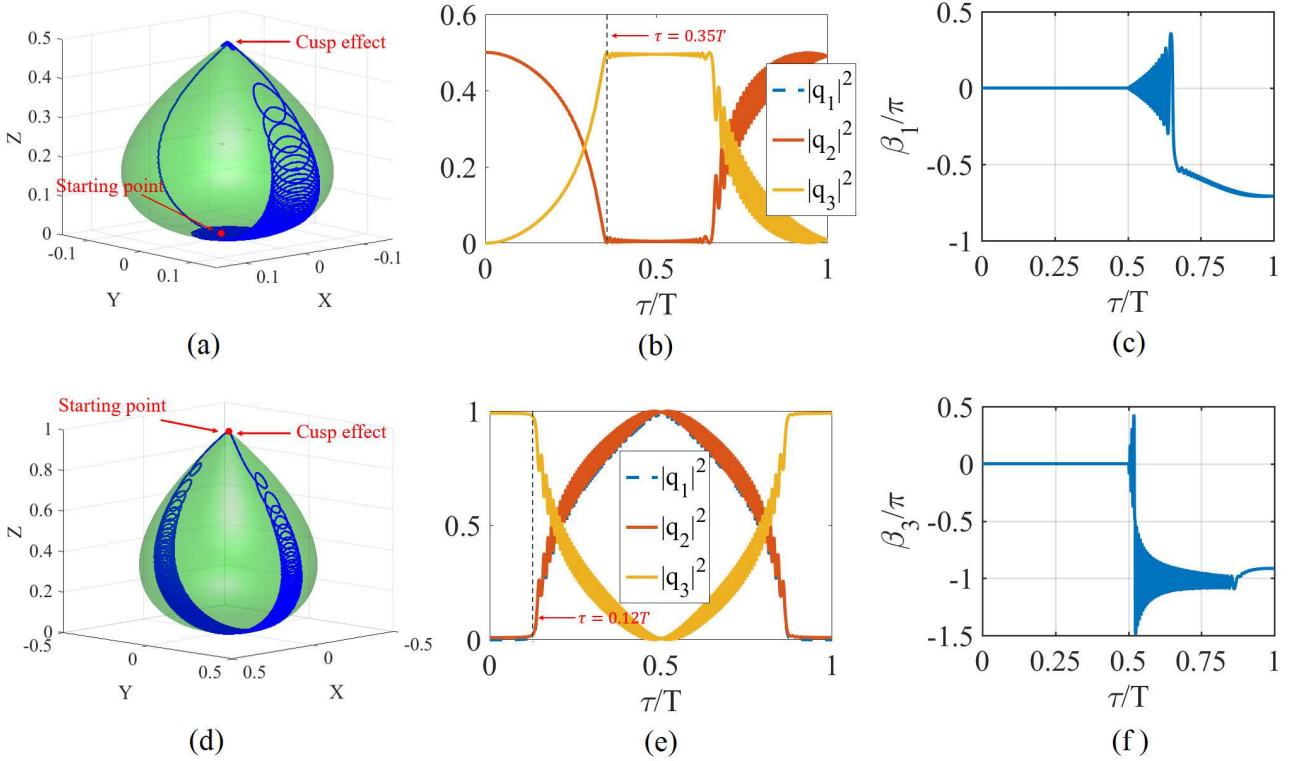}
\caption{Cusp effect for the full-wedge rotation in the SFG (a,b c) and DFG (d,e,f) processes. (a) Bloch surface and trajectory for SFG with $I_{1}=I_{2}$. (b) Evolution of the intensities of three waves. At $\tau=0.35T$, the state vector $\mathbf{W}$ reaches the cusp. (c) Corresponding geometric phase accumulation of $\beta_{1}$ in panel (a). (d) Bloch surface and trajectory for DFG with $I_{2}=0.01$ and $I_{3}=0.99$. (e) Evolution of the intensities of three waves. At $\tau=0.12T$, the state vector $\mathbf{W}$ leaves the cusp. (f) Corresponding geometric phase accumulation of $\beta_{3}$ in panel (c). Here, in all panels, we select $\Delta\phi_{d}=\pi$.} \label{Cusp}
\end{figure}

We first consider the case of SFG with different depletion levels. In this case, the initial condition satisfies
\begin{eqnarray}
I_{1,2}\neq0, I_{3}=0.
\end{eqnarray}
The dynamics of state vector $\mathbf{W}$ on the Bloch surface start from the bottom of the surface. The full-wedge rotation of $\mathbf{Q}$ makes the state vector $\mathbf{W}$ draw a full-wedge-shaped trajectory on the surface. Hence, this trajectory may pass the north pole of the Bloch surface. A typical example of this process is shown in Fig. \ref{Fwedgeresult}(a1). When the initial intensity of the idler is smaller than that of the pump, the geometric phase solely accumulates for the idler wave even though the pump wave is depleted. The geometric phase of the idler wave is $\beta_{1}=-\Delta\phi_\mathrm{d}$, which is the same as that in the undepleted pump case (i.e., $I_{1}\ll I_{2}$). A typical example of geometric phase accumulation of $\beta_{1}$ and $\beta_{2}$ with $I_{1}:I_{2}=4:6$ is displayed in Fig. \ref{Fwedgeresult}(a2). If the case is switched to $I_{1}>I_{2}$, then the geometric phase accumulation is also switched to $I_{2}$, which has a symmetric relationship with the case of $I_{1}<I_{2}$. At $I_{1}=I_{2}$, the system is at the full depletion limit, and a sharp cusp appears at the top of the Bloch surface [see Fig. \ref{Cusp}(a)]. Fig. \ref{Cusp}(b) displays the evolution of the intensities of three waves at the case of the full depletion limit with $\Delta\phi_{\mathrm{d}}=\pi$. At this case the intensity of $\omega_{1}$ and $\omega_{2}$ are completely identical to each other, which results in the completely overlap between $|q_{1}|^2$ and $|q_{2}|^2$ in Fig. \ref{Cusp}(b). At $\tau=0.35T$, $q_{1}=q_{2}=0$, which gives rise to $X=Y=0$ [See the definition in Eq. (\ref{coordinate})]. Because $q_{3}\neq0$ (i.e., $Z\neq0$), the state vector $\mathbf{W}$ reaches the cusp at $\tau=0.35T$. According to full-wedge motion of QPM vector $\mathbf{Q}$ in Section 2.3, the vector $\mathbf{Q}$ reaches the north pole of the parameter surface at $\tau=0.5T$. This truth indicates that vector $\mathbf{W}$ and vector $\mathbf{Q}$ do not move together when they are close to the cusp. Once vector $\mathbf{W}$ does not move together with vector $\mathbf{Q}$, $\mathbf{W}$ no longer represent the eigenstate of the system, which gives rise to the failure accumulation of the geometric phase [see Fig. \ref{Cusp}(c)]. Fig. \ref{Fwedgeresult}(a3) shows $\beta_1$ versus $\Delta\phi_d$ at different pump depletion levels. For convenience, here, we consider only the case of $I_{1}<I_{2}$ (geometric phase is accumulated only on $\beta_{1}$). The figure shows that except for at the full depletion limit, the geometric phase remains constant when QPM vector $\mathbf{Q}$ undergoes full-wedge rotation. Under this circumstance, the geometric phase is independent of the pump depletion level.

For the case of DFG, the initial condition satisfies
\begin{eqnarray}
I_{2,3}\neq0, I_{1}=0.
\end{eqnarray}
In this case, the dynamics of state vector $\mathbf{W}$ on the Bloch surface start from the top of the surface, and the geometric phase accumulates only on $\beta_{3}$. When the intensity of the signal ($q_{3}$) becomes comparable to that of the pump ($q_{2}$), the pump wave becomes depleted. A typical example of the trajectory of the state vector on the Bloch surface and the geometric phase accumulations of $\beta_{3}$ and $\beta_{2}$ are displayed in Fig. \ref{Fwedgeresult}(b1,b2). Full depletion occurs at the limit of $I_{3}\rightarrow1$ and $I_{2}\rightarrow0$; under this condition, a sharp cusp also appears at the top of the Bloch surface [see Fig. \ref{Cusp}(d)]. Fig. \ref{Cusp}(e) displays the evolution of intensities of the 3 waves for this case also with $\Delta\phi_{\mathrm{d}}=\pi$. This figure shows that the state vector $\mathbf{W}$ starts to leave the cusp at $\tau=0.12T$, while the vector $\mathrm{Q}$ starts to leave the north pole at $\tau=0$. This truth indicates that $\mathbf{W}$ does not move together with vector $\mathbf{Q}$ at the beginning of the dynamics. This cusp effect also destroys the geometric phase accumulation [see Fig. \ref{Cusp}(f)]. Similar to the case of SFG, except for at the full depletion limit, the geometric phase is independent of the pump depletion level in the case of DFG when the QPM vector undergoes full-wedge rotation. Fig. \ref{Fwedgeresult}(b3) displays the geometric phase $\beta_{3}$ versus $\Delta\phi_d$ for different pump depletion levels. This result indicates that the magnitude of the accumulated geometric phase is identical to the case of the undepleted pump limit.

Therefore, we can conclude that when the QPM vector undergoes full-wedge rotation, the geometric phase is independent of the pump depletion level. The magnitude of the geometric phase is identical to that at the undepleted pump limit. This result is different from that in the case when the QPM vector undergoes circular rotation, which shows dependence of the geometric phase and the pump depletion level. In the case of circular rotation, all three QPM parameters vary adiabatically, while in the case of full-wedge rotation, the phase factor of QPM parameter $\phi_d$ exhibits a nonadiabatic variation. Hence, it seems that the adiabatic variation in the phase factor makes the geometric phase depend on the pump depletion level. To demonstrate this statement, we consider half-wedge rotation of the QPM vector, in which $\phi_d$ undergoes adiabatic variation.

\subsection{Half-wedge rotation}

\begin{figure}[htbp]
\centering\includegraphics[width=0.66\columnwidth]{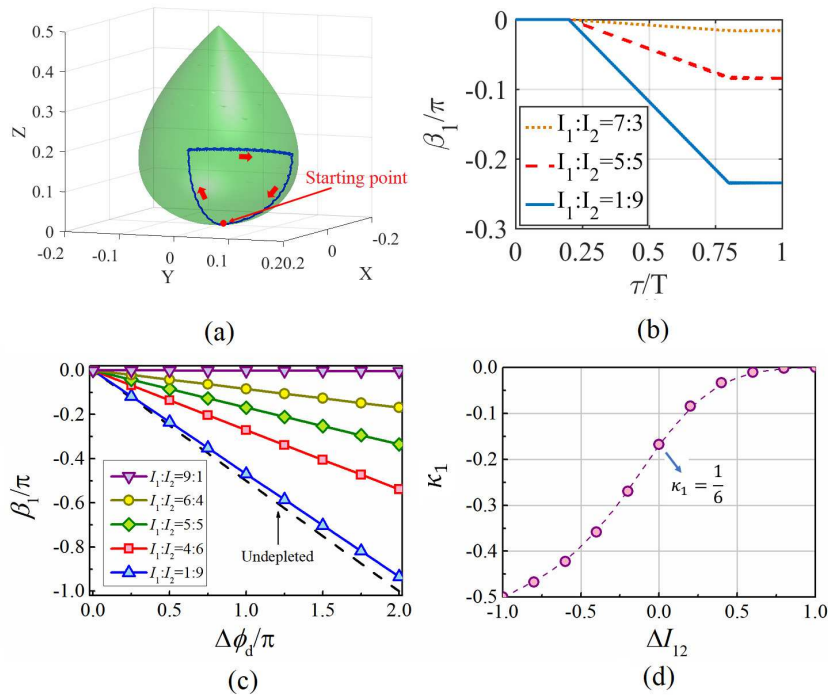}
\caption{(a) Trajectory of the clockwise half-wedge rotation (blue curve) in the SFG process on the Bloch surface with $I_{1}:I_{2}=5:5$ and $\Delta\phi_d=\pi/2$. This process starts from the bottom of the Bloch surface. (b) Accumulation of $\beta_{1}$ for different pump depletion levels. (c) $\beta_{1}$ (at the output end) versus $\Delta\phi_d$ for different depletion levels. At the undepleted limit, $\beta_{1}=-\Delta\phi_d/2$ (the black dashed line). (d) $\kappa$, the proportionality coefficient between $\beta_{1}$ and $\Delta\phi_{d}$, versus $\Delta I_{12}=I_{1}-I_{2}$. At the case of $I_{1}=I_{2}$ (i.e., $I_{12}=0$), $\kappa_{1}=1/6$. } \label{Hwedgeresult}
\end{figure}
\begin{figure}[htbp]
\centering\includegraphics[width=0.66\columnwidth]{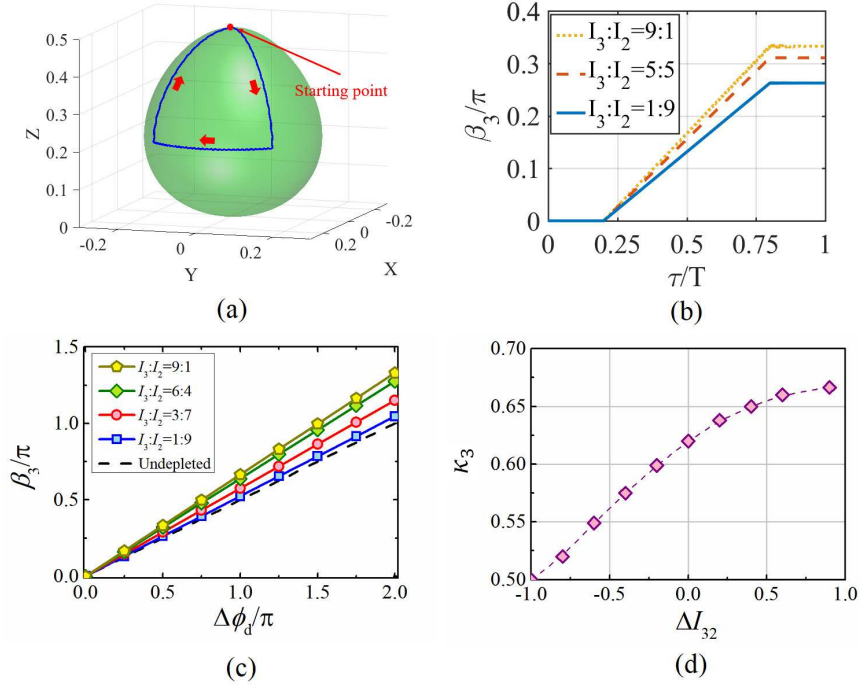}
\caption{(a) Trajectory of the clockwise half-wedge rotation (blue curve) in the DFG process on the Bloch surface with $I_{3}:I_{2}=5:5$ and $\Delta\phi_d=\pi/2$. This process starts from the top of the Bloch surface. (b) Accumulation of $\beta_{3}$ for different pump depletion levels. (c) $\beta_{3}$ (at the output end) versus $\Delta\phi_d$ for different depletion levels. At the undepleted limit, $\beta_{3}=\Delta\phi_d/2$ (the black dashed line). (d) $\kappa$, the proportionality coefficient between $\beta_{3}$ and $\Delta\phi_{d}$, versus $\Delta I_{32}=I_{3}-I_{2}$.} \label{Hwedgeresult2}
\end{figure}

An example of a trajectory of $\mathbf{W}$ for half-wedge rotation on the Bloch surface is displayed in Fig. \ref{Hwedgeresult}(a). Fig. \ref{Hwedgeresult}(b) shows the geometric phase accumulation of $\beta_{1}$ for different pump depletion levels. As expected, the geometric phase accumulation depends on the pump depletion level when $\phi_{d}$ undergoes an adiabatic variation during rotation. Fig. \ref{Hwedgeresult}(c) shows $\beta_{1}$ versus the wedge angle $\Delta\phi_{d}$ for different pump depletion levels. Interestingly, $\beta_{1}$ satisfies
\begin{eqnarray}
\beta_{1}=\kappa_{1}\Delta\phi_{d},
\end{eqnarray}
where $\kappa_{1}$ as a function of $\Delta I_{12}=I_{1}-I_{2}$ is displayed in Fig. \ref{Hwedgeresult}(d). We also consider the case of half-wedge rotation in the DFG process, which starts its dynamics from the top of the Bloch surface. Similar to the SFG case, geometric phase $\beta_{3}$ in this case becomes dependent on the pump depletion level and satisfies $\beta_{3}=\kappa_{3}\Delta\phi_{d}$. A typical example of the trajectory of $\mathbf{W}$ on the Bloch surface, the accumulation of $\beta_{3}$ for different pump depletion levels, $\beta_{3}$ versus wedge angle $\Delta\phi_{d}$, and $\kappa$ versus $\Delta I_{32}=I_{3}-I_{2}$ are displayed in Fig. \ref{Hwedgeresult2}. Unlike in the SFG process, in the case of the fully depleted pump limit in the DFG process, i.e., $I_{3}\rightarrow1$ and $I_{2}\rightarrow0$, the cusp effect is induced at the starting point, and $\beta_{3}$ cannot accumulate in this case. Hence, the interval for $\Delta I_{32}$ is $[-1,1)$. In contrast, for the SFG case, because the trajectory of $\mathbf{W}$ does not need to pass the top of the Bloch surface, $\beta_{1}$ can be successfully accumulated at the fully depleted pump limit (i.e., the case of $I_{1}=I_{2}$).

Even though the geometric phase becomes dependent on the pump depletion level in the case of half-wedge rotation, the linear dependence between the geometric phase and the wedge angle, i.e., $\beta_{1}=\kappa_{1}\Delta\phi_{d}$ in the SFG process and $\beta_{3}=\kappa_{3}\Delta\phi_{d}$ in the DFG process, indicates that such a dependence shows a better controllability than that in the case of circular rotation in Ref. \cite{31:Yongyao}.

\section{Application}

\begin{figure}[htbp]
\centering\includegraphics[width=0.66\columnwidth]{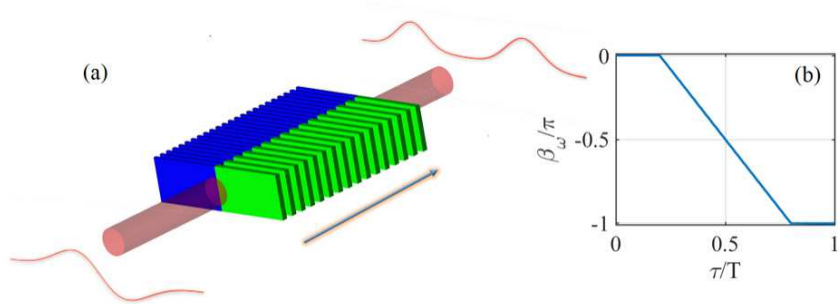}
\caption{(a) The crystal is divided by two adiabatic QPM paths. One path is assumed to be designed by a half-wedge rotation with $\Delta\phi_d=6\pi$ (i.e., the $\mathbf{Q}$ vector rotates for three rounds at the equator, which overall creates a $\pi$ geometric phase shift). The other path is a corresponding three rounds of round-trip motion, which creates the same dynamical phase shift as the half-wedge rotation. An HG01-like mode is input to this crystal, with two humps transmitting in the two paths. Because the $\pi$-phase difference between the two humps is compensated by the geometric phase, an in-phase two-hump structure is created at the output end. (b) Geometric phase accumulation process for the half-wedge rotation with $\Delta\phi_d=6\pi$.} \label{SFGscheme}
\end{figure}

The geometric phase in the case of full-wedge rotation is stable even when the pump becomes depleted, whereas it cannot be applied to the fully depleted pump level case [see Fig. \ref{Cusp}]. In other words, the full-wedge rotation of the QPM vector cannot be applied to control the geometric phase in the case of SHG, which satisfies $\omega_{1}=\omega_{2}=\omega$, and $\omega_{3}=\omega_{1}+\omega_{2}=2\omega$, at which the condition $I_{1}=I_{2}$ is automatically valid because $\omega_{1}$ and $\omega_{2}$ are mixed into a single light field. At the case of SHG, Eqs. (\ref{Basieq1},\ref{Basieq2}) are degenerated to a single equation. However, if we adopt the transformations
\begin{eqnarray}
&&A_{1}=A_{2}=A_{\omega}/\sqrt{2}, A_{3}=A_{2\omega},\nonumber\\
&&n_{1}=n_{2}=n_{\omega}, n_{3}=n_{2\omega},\nonumber\\
&&\omega_{1}=\omega_{2}=\omega, \omega_{3}=2\omega,
\end{eqnarray}
we can still use the configuration of 3 equations to simulate the process of SHG. As discussed above, the cusp effect makes full-wedge rotation fail to create the geometric phase. However, for the case of half-wedge rotation, the trajectory of the state vector does not need to pass the cusp of the Bloch surface, and the cusp effect can be avoided. Hence, the half-wedge rotation provides a possibility to arbitrarily control the geometric phase generation for the light of the first harmonic in the SHG process.

Fig. \ref{SFGscheme}(a) shows a sketch map for transforming a Hermite-Gaussian-01 (HG01)-like mode to an in-phase two-hump mode during the SHG process by means of half-wedge rotation of the QPM in nonlinear crystals. The crystal is divided into two parts. The QPM vector in the first part is rotated by a half-wedge rotation with $\Delta\phi_{d}=6\pi$. According to the numerical results in Fig. \ref{Hwedgeresult}(d), $\kappa_{1}=1/6$ at $I_1=I_2$), the geometric phase for the first harmonic is $\beta_{\omega}=-\pi$. The corresponding geometric phase accumulation process is shown in Fig. \ref{SFGscheme}(b). Therefore, the total phase difference at the output end is
\begin{eqnarray}
\Phi^{(\mathrm{I})}_{\omega}=D_{\omega}-\pi,
\end{eqnarray}
where $D_{\omega}$ and $\pi$ are the dynamical phase and the geometric phase, respectively. However, the QPM vector in the next part undergoes only round-trip motion, which varies only $\Delta\Gamma$ and $\Xi$. $\phi_{d}\equiv0$ for the entire process. Round-trip motion accumulates only a dynamical phase and no geometric phase. Hence, the total phase difference in this part is
\begin{eqnarray}
\Phi^{(\mathrm{II})}_{\omega}=D_{\omega}.
\end{eqnarray}
Therefore, the phase difference between these two parts is $\Delta\Phi_{\omega}=\Phi^{(\mathrm{I})}_{\omega}-\Phi^{(\mathrm{II})}_{\omega}=-\pi$. This phase shift compensates for the $\pi$ phase shift difference between the two humps of the HG01-like mode and transforms it into an in-phase two-hump mode. In principle, when a Gaussian beam is injected into the crystal, the diffractive term should be involved into Eq. (\ref{Basieq1}-\ref{Basieq3}). However, if the width of the Gaussian beam is broad enough, which results in the diffractive length of beam being much longer than the length of the crystal, one can neglect the diffractive term from the equation. The CW-equation may still be valid for this case. Recently, a similar experimental setting for this manipulation was realized in Ref. \cite{26:Aviv}, but this setting requires inputting a strong pump to achieve this target.

\section{Conclusion}

In summary, the objective of this work is to create or manipulate a well-predicted geometric phase for idler or signal waves in fully nonlinear three-wave mixing via the SFG or DFG process with different pump depletion levels. A previous study showed that the magnitude of the accumulated geometric phase strongly depends on the pump depletion level. In this paper, we found that the adiabatic variation and nonadiabatic variation in the QPM parameters play an important role in the relationship between the geometric phase and the pump depletion level. This finding is based on the geometric phase accumulation in two rotation schemes, full-wedge rotation and half-wedge rotation, of the QPM parameters in the process of nonlinear three-wave mixing. In the case of full-wedge rotation, the phase factor of the QPM parameter varies nonadiabatically during rotation. A step function for the phase factor occurs when the QPM vector reaches the pole of the surface in the parameter space. In this case, the geometric phase is independent of the pump depletion level. Its magnitude remains constant and equal to that at the undepleted pump limit. This result indicates that we can create a certain geometric phase by using this configuration even when the pump intensity becomes unstable. However, for the case of half-wedge rotation, when all QPM parameters vary adiabatically during the rotation, the geometric phase becomes dependent on the pump depletion level. In contrast to the previous study on circular rotation, the geometric phases at each pump depletion level are all linearly dependent on the wedge angle, showing much more predictable properties than those in circular rotation. Finally, we provide an example of a potential application for all-optical transformation of the HG-01-like mode into an in-phase two-hump mode by a half-wedge rotation with $\Delta\phi_{d}=6\pi$ in the SHG process.

\begin{acknowledgments}
Y. L. appreciates the useful discussion from Ofir Yesharim (Tel Aviv University), Aviv Karnieli
(Tel Aviv University), Prof. Ady Arie (Tel Aviv University), and Prof. Gil Porat (University of Alberta). This work was supported by the NNSFC (China) through Grant Nos. 11874112 \& 62005044, the Project of Department of Education of Guangdong Province (No. 2018KTSCX241), and the Research Fund of Guangdong-Hong Kong-Macao Joint Laboratory for Intelligent Micro-Nano Optoelectronic Technology (No. 2020B1212030010).
\end{acknowledgments}

\end{document}